\documentclass[12pt]{article}

\usepackage{graphicx,psfrag,epsf,color}
\usepackage{amsmath,amssymb,amsfonts}
\usepackage{array}
\usepackage{cite}
\usepackage{multirow}

\renewcommand{\arraystretch}{1.25}

\newcounter{MBQ}

\newcommand{\braket}[1]{\langle #1 \rangle}
\newcommand{\Tprod}[1]{{\mathrm T}\lbrack #1 \rbrack}
\begin{document}

\allowdisplaybreaks

\begin{titlepage}

\begin{flushright}
{\small
TUM-HEP-1045/16\\
%1606.nnnnn [hep-ph]\\[0.0cm]
June 7, 2016}
\end{flushright}

\vskip1cm
\begin{center}
{\Large\bf\boldmath
Threshold singularities, dispersion relations and\\[0.1cm] 
fixed-order perturbative calculations
}
\end{center}

\vspace{0.8cm}
\begin{center}
{\sc M.~Beneke} and {\sc P. Ruiz-Femen\'\i a}\\[5mm]
{\it Physik Department T31,\\
James-Franck-Stra\ss e, 
Technische Universit\"at M\"unchen,\\
D--85748 Garching, Germany
}
\end{center}

\vspace{1cm}
\begin{abstract}
\vskip0.2cm\noindent
We show how to correctly treat threshold singularities in fixed-order 
perturbative calculations of the electron anomalous magnetic moment 
and hadronic pair production processes such as top pair production. 
With respect to the former, 
we demonstrate the equivalence of the ``non-perturbative'', resummed 
treatment of the vacuum polarization contribution, whose spectral function 
exhibits bound state poles, with the fixed-order calculation by 
identifying a threshold localized term in the four-loop spectral 
function. In general, we find that a modification of the dispersion 
relation by threshold subtractions is required to make fixed-order 
calculations well-defined and provide the subtraction term. We then solve the 
apparent problem of a divergent convolution of the partonic cross section 
with the parton luminosity in the computation of the top pair production 
cross section starting from the fourth-order correction. We find 
that when the computation is performed in the usual way as 
an integral of real and virtual corrections over phase space at a given 
order in the expansion in the strong coupling, 
an additional contribution has to be added at N3LO. 
\end{abstract}
\end{titlepage}

%\setcounter{page}{1}

%%%%%%%%%%%%%%%%%%%%%%%%%%%%%%%%%%%%%%%%%%%%%%%%%%%%%%%%%%%%%%%%%%%%%%%%%%%
\section{Introduction}
\label{sec:intro}

\begin{figure}[t]
\begin{center}
\includegraphics[scale=0.75]{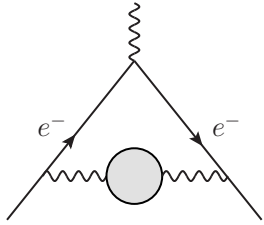}
\caption{Vacuum polarization contribution to the anomalous magnetic 
moment of the electron.}
\label{fig:vpdiag}
\end{center}
\end{figure}

The photon vacuum polarization contribution to the electron anomalous 
magnetic moment $g_e-2$, see Figure~\ref{fig:vpdiag} is 
given by~\cite{Lautrup:1969uk,Lautrup:1971jf}
\begin{align}
a_e^{\rm (vp)} & = - \frac{\alpha}{\pi} \int_0^1 dx \,(1-x) \, 
\Pi\Big(\frac{-x^2}{1-x}\,m^2\Big)
\,,
\label{aevp}
\end{align}
with $\alpha$ the fine structure constant and $m$ the electron mass.
Exploiting the analyticity of $\Pi(s)$ and the standard on-shell 
renormalization condition $\Pi(0)=0$, the once-subtracted dispersion relation 
\begin{align}
\Pi(q^2) &=  \frac{q^2}{2\pi i} \oint ds \, \frac{\Pi(s)}{s\,(s-q^2)} 
  = \frac{q^2}{\pi}\int_0^\infty \frac{ds}{s} \,
  \frac{ {\rm Im}\,\Pi(s+i\eta) }{s-q^2}
\label{disprel}
\end{align}
holds, which allows us to rewrite (\ref{aevp}) as
\begin{align}
a_e^{\rm (vp)} & =  \frac{\alpha}{\pi^2} 
\int_0^\infty \frac{ds}{s} \,{\rm Im}\,\Pi(s+i\eta)\,K(s) 
\label{dispaevp}
\end{align}
with kernel function 
%\vskip-0.7cm
\begin{align}
K(s) & =  \int_0^1 dx \, \frac{x^2(1-x)}{x^2+(1-x)s/m^2} 
\;.
\label{K}
\end{align}

The spectral function ${\rm Im}\,\Pi(s)$ exhibits a series of 
positronium poles\footnote{The spectral function is often discussed in 
connection with hadronic contributions to the anomalous magnetic 
moment. Here we are concerned with QED effects only. We also note that 
in QED the discontinuity of $\Pi(s)$ starting at $s=0$ 
is due to three-photon intermediate states, which first enter at 
${\cal O}(\alpha^4)$ in the perturbative vacuum polarization. For the 
purposes of this paper, we are only interested
in the $e^+e^-$ physical cut, which starts at $s=4m^2$, and 
the positronium poles slightly below.}
slightly below the electron-positron threshold $4 m^2$. In 
\cite{Mishima:2013ama} it has been claimed that this results 
in an additional ${\cal O}(\alpha^5)$ contribution to the magnetic moment, 
which is not captured by the ${\cal O}(\alpha^5)$  QED correction from 
the four-loop vacuum polarization function~\cite{Aoyama:2010zp}.  This 
claim has been quickly 
refuted \cite{Melnikov:2014lwa,Eides:2014swa,Hayakawa:2014tla} --- 
indeed, it is clear from (\ref{aevp}) that the vacuum polarization 
is probed only in the Euclidean region far from the 
electron-positron threshold, where an ordinary loop expansion is 
valid ---, but the arguments presented leave an interesting point open, 
namely whether and how an order-by-order calculation of the 
spectral function gives the correct result for the magnetic 
moment when the dispersive representation (\ref{dispaevp}) is 
used. The answer to this question, which we provide in this note, 
leads to more general considerations on the formulation of the 
dispersion relation for spectral functions whose perturbative 
expansions become more and more singular near pair-production thresholds 
as the order of the expansion increases. This in turn has interesting 
ramifications for pair production of heavy particles such as top quarks 
at the Large Hadron Collider as will be discussed.

%%%%%%%%%%%%%%%%%%%%%%%%%%%%%%%%%%%%%%%%%%%%%%%%%%%%%%%%%%%%%%%%%%%%%%%%
%%%%%%%%%%%%%%%%%%%%%%%%%%%%%%%%%%%%%%%%%%%%%%%%%%%%%%%%%%%%%%%%%%%%%%%%

\section{ \boldmath Equivalence at $ {\cal O}(\alpha^5)$}
\label{sec:equivalence}

The vacuum polarization function develops poles of $e^+e^-$ bound states 
right below the electron-positron threshold. They cannot 
be obtained at any finite order in QED perturbation theory, but arise 
diagrammatically from the summation of an infinite number of Coulomb-photon
exchanges between the electron and positron. The systematic resummation 
can be performed within the framework of non-relativistic effective 
field theory, and the relevant counting is $v\equiv\sqrt{E/m} \sim \alpha$ 
with $E=\sqrt{s}-2 m$.

The summation generates the bound-state poles and also significantly 
affects the $e^+ e^-$ continuum near threshold. For the present 
discussion of ${\cal O}(\alpha^5)$ effects both are adequately 
described by the leading-order Coulomb Green function. The photon 
vacuum polarization near threshold (small $E$) is given by
\begin{align}
\Pi_0(E) & = \frac{2\pi\alpha}{m^2}\,G_0(0,0;E)
\label{eq:PiGrelation}
\end{align}
in terms of the zero-distance Coulomb Green function 
\cite{Eiras:1999xx,Beneke:1999zr}
\begin{align}
G_0(0,0;E) & =
\frac{m^2}{4\pi}\left[-\sqrt{-\frac{E}{m}} -\alpha
\left\{-\frac{1}{4\epsilon} 
+\frac{1}{2}\ln\left(\frac{-4 m E}{\mu^2}\right) -\frac{1}{2} 
+\gamma_E+\Psi(1-\lambda)
\right\}\right],
\label{eq:G0MSbar}
\end{align}
here regulated dimensionally in $d=4-2\epsilon$ dimensions.
The Coulomb Green function sums terms of order $(\alpha/v)^n$ to 
all orders in $\alpha$ through the digamma function $\Psi(1-\lambda)$,
where $\lambda=\alpha/(2\sqrt{-E/m})$, and the poles of the digamma function
at positive integer $\lambda$ correspond to the $S$-wave positronium 
bound states. The imaginary part of the Green function for real energies 
reads
\begin{align}
{\rm Im}\,G_0(0,0;E) & =
\sum_{n=1}^\infty \left(\frac{ m\alpha}{2n}\right)^3
\, \delta(E-E_n) + \theta(E)\, \frac{m^2}{4\pi} \,
\frac{\pi \alpha}{1-e^{-\frac{\pi\alpha}{v}}}\,,
\label{eq:ImG0}
\end{align}
where the second term is the continuum contribution known as the Sommerfeld 
factor and the positronium bound states at energies $E_n = 
-m\alpha^2/(4 n^2)$ are explicit in the first term. We shall now 
compute the contribution to the anomalous magnetic moment in two ways. 
First, ``non-perturbative'', that is, using the all-order resummed spectral 
function above. Second, we show that exactly the same result can be 
obtained at fixed five-loop order in perturbation theory. We then explain 
why this implies that no additional contribution has to be added to 
the known result~\cite{Aoyama:2010zp}. 

For the ``non-perturbative'', resummed evaluation we multiply (\ref{eq:ImG0}) 
by $2\pi\alpha/m^2$ and insert the result into (\ref{dispaevp}), obtaining 
\begin{align}
a_e^{\rm (vp), res} &=
\frac{\alpha^5}{4\pi} \, \sum_{n=1}^\infty \, \frac{1}{n^3} \, 
\frac{K((2m+E_n)^2)}{1+E_n/2m}
+ \, \frac{\alpha^3}{\pi} \int_0^\infty \!\!\! dE \, 
\frac{K((2m+E)^2)}{2m+E}
\frac{1}{1-e^{-\frac{\pi\alpha}{\sqrt{E/m}}}}\,.
\label{eq:resummed}
\end{align}
The first term represents the positronium contribution. At  
$ {\cal O}(\alpha^5)$ we can neglect the $E_n/m$ corrections, and 
the expression evaluates to $\alpha^5\zeta_3/(4\pi)\times K(4 m^2)$, 
where $\zeta_k=\sum_{n=1}^\infty 1/n^k$ is the Riemann zeta function.
The integral over the continuum spectral function contains lower 
order contributions starting from $ {\cal O}(\alpha^2)$, which are 
of no interest here, and is divergent at large $E$, which is an 
artifact, since the employed approximation to ${\rm Im}\,G_0$ applies
only for small $E\ll m$. Subtracting the lower order contributions and 
applying a cut-off $E_{\rm max} = m v_{\rm max}^2$ to the energy 
integral, we are left with 
\begin{align}
\frac{\alpha^3}{\pi}\,K(4 m^2)\int_0^{v_{\rm max}} \!\!\! dv\,v
\left( \frac{1}{1-e^{-\frac{\pi\alpha}{v}}} - 
        \frac{v}{\pi\alpha}-\frac{1}{2}-\frac{\pi\alpha}{12v}
\right)
\label{eq:contint}
\end{align}
Note that we are allowed and must choose $v_{\rm max}$ such 
that $\alpha \ll v_{\rm max}\ll 1$ in order to include the non-perturbative 
modification of the threshold region. It is straightforward to check that 
the largest contribution to the integral is ${\cal O}(\alpha^2)$ and 
arises from the region $v\sim \alpha$, while for $v\gg \alpha$ 
the integrand behaves as $\alpha^3/v^2$ and hence the contribution from 
that region is at most of order $\alpha^3/v_{\rm max}\ll \alpha^2$. This 
allows us to set the upper integration limit $v_{\rm max}$ to infinity 
and to obtain the analytic result  $-\alpha^5\zeta_3/(8\pi)\times K(4 m^2)$
(already given in \cite{Melnikov:2014lwa}) for the above expression 
(\ref{eq:contint}).\footnote{An even simpler way to obtain this 
result, which can be justified in the context of the threshold 
expansion \cite{Beneke:1997zp}, is to apply an analytic regulator 
$v\to v^{1+\lambda}$ to the integrand factor in (\ref{eq:contint}) 
and to compute
\[\int_0^{\infty} \!\!\! dv\,v^{1+\lambda}
\frac{1}{1-e^{-\frac{\pi\alpha}{v}}} = -\frac{\alpha^2\zeta_3}{8} 
+{\cal O}(\lambda)\,,
\]
which extracts the contribution from $v\sim\alpha$, which cannot be 
obtained from the Taylor expansion of the integrand in $\alpha$.}
Thus, the threshold contribution to the anomalous magnetic moment 
from the fourth-order vacuum polarization is 
\begin{align}
\big[ a_e^{\rm (vp), res}\big]_{{\cal O}(\alpha^5)} & =
\underbrace{\frac{\alpha^5\zeta_3}{4\pi}\,K(4m^2)}_{\rm positronium\,\,poles}-
\underbrace{\frac{\alpha^5\zeta_3}{8\pi}\,K(4m^2)}_{\rm continuum}\,.
\label{eq:resummed2}
\end{align}

We now turn to the second, perturbative evaluation. The expansion 
of $\Pi(E)$ in $\alpha$ can be recovered by expanding $G_0(0,0;E)$ in 
$\alpha$, since the threshold approximation is sufficient for the present 
purpose. Up to the four-loop order, we find
\begin{align}
\Pi_0^{\rm pert}(E)  = & 
-\frac{\alpha}{2}\,\sqrt{\frac{-E}{m}}
-\frac{\alpha^2}{4}\bigg( -\frac{1}{2\epsilon}+
\ln \Big(\frac{-4m E}{\mu^2}\Big) -1 \bigg)
+\frac{\pi^2\alpha^3}{24}\frac{1}{\sqrt{-E/m}}
\nonumber\\
&-\frac{\alpha^4}{8}\,\frac{\zeta_3}{E/m}
+{\cal O}(\alpha^5)\,.
\label{eq:Pipert}
\end{align}
The ${\cal O}(\alpha^4)$ contribution proportional to $1/E$ 
is particularly relevant for the present discussion.\footnote{This term 
can also be identified from the most singular term in the threshold 
expansion of the full four-loop vacuum polarization given in the 
appendix of~\cite{Kiyo:2009gb}.} Interpreting 
$\Pi(E)$ as a distribution with $E\to E+i \eta$, where $\eta$ is 
positive-infinitesimal, this term implies a threshold-localized 
contribution to the four-loop spectral function given by
\begin{align}
\big[ {\rm Im}\,\Pi_0^{\rm pert}(E+i\eta) \big]_{{\cal O}(\alpha^4)}  = &
\; \frac{\pi\alpha^4\zeta_3}{8}\,m\delta(E) \,.
\label{eq:ImPipert}
\end{align}
Using this in (\ref{dispaevp}) we find that the threshold contribution 
to the anomalous magnetic moment from the fourth-order vacuum polarization 
is 
\begin{align}
\big[ a_e^{\rm (vp), pert}\big]_{{\cal O}(\alpha^5)} & =
\frac{\alpha^5\zeta_3}{8\pi}\,K(4m^2)\,,
\label{eq:F)}
\end{align}
in precise agreement with (\ref{eq:resummed2}). Thus we have shown 
that the fixed-order perturbative approximation accurately reproduces the 
threshold contribution of the exact spectral function {\em including the 
positronium pole contribution, provided the threshold singularities 
of the vacuum polarization are interpreted in the distribution sense.}

Let us add the following remarks. 1) The ``non-perturbative'' evaluation 
essentially coincides with the derivation in~\cite{Melnikov:2014lwa}, 
but the perturbative one is different, since the $\delta(E)$ term in 
${\rm Im}\,\Pi_0^{\rm pert}$ was not identified there. 
Instead, analyticity was invoked to 
relate the energy integral over ${\rm Im}\,G_0(0,0;E)$, which appears in 
$a_e^{\rm (vp)}$, to the asymptotic behaviour of $E G_0(0,0;E)$ at 
$E\to -\infty$. This step, while mathematically correct, is nevertheless 
physically somewhat dubious, since it should not be necessary to appeal to 
the behaviour of the Coulomb Green function outside its range of 
applicability. 2) The direct expansion in $\alpha$ 
of ${\rm Im}\,G_0$ in (\ref{eq:ImG0}) using $\delta(E-E_n)= 
\delta(E) + {\cal O}(\alpha^2)$ would yield the wrong result, namely
\begin{align}
\Big[ \frac{2\pi\alpha}{m^2} \, 
{\rm Im}\,G_0^{\rm pert}(0,0;E)\Big]_{{\cal O}(\alpha^4)} & =
\frac{\pi \alpha^4\zeta_3}{4} \, m \delta(E)\,,
\label{eq:ImG0pert}
\end{align}
which differs from the correct result (\ref{eq:ImPipert}) by a factor of 
two. One must either integrate the resummed spectral function properly 
{\em or} derive the perturbative spectral function from the expansion 
of $\Pi_0(E)$ in the distribution sense. 3) Although the threshold 
contribution (\ref{eq:resummed2}) or (\ref{eq:F)}) is non-zero, this 
does {\em not} imply that it has to be added to the result 
of~\cite{Aoyama:2010zp}. In this paper $\Pi(s)$ is computed directly, and 
(\ref{aevp}) is employed to obtain the anomalous magnetic moment, hence 
the subtlety of the threshold-localized $\delta(E)$ in term in 
the four-loop spectral function never arises.

%%%%%%%%%%%%%%%%%%%%%%%%%%%%%%%%%%%%%%%%%%%%%%%%%%%%%%%%%%%%%%%%%%%%%%%%%%%%
\section{Threshold-subtracted dispersion relations}
\label{sec:thrsing}

%%%%%%%%%%%%%%%%%%%%%%%%%%%%%%%%%%%%%%%%%%%%%%%%%%%%%%%%%%%%%%%%%%%%%%%%
\begin{figure}[t]
\begin{center}
\includegraphics[scale=0.55]{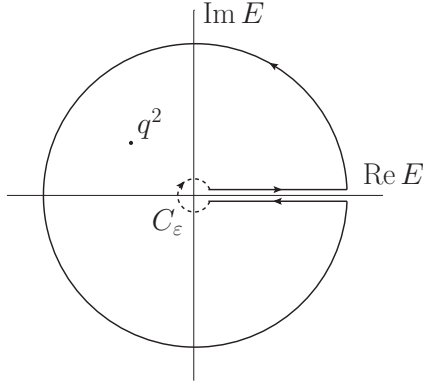}
\caption{Integration contour defining the dispersion relation for 
$\Pi(E)$. }
\label{fig:intcircuit}
\end{center}
\end{figure}
%%%%%%%%%%%%%%%%%%%%%%%%%%%%%%%%%%%%%%%%%%%%%%%%%%%%%%%%%%%%%%%%%%%%%%%%

The $n$-loop vacuum polarization behaves as $\Pi^{(n)}(s)\propto 
\alpha^{n} E^{1-n/2}$ when $E=\sqrt{s}-2m \to 0$. The existence of 
non-integrable singularities at the $e^+e^-$ threshold starting at 
${\cal O}(\alpha^4)$ calls for a careful analysis of the dispersion 
relation (\ref{disprel}). The first equality in (\ref{disprel}) holds 
order by order, hence 
\begin{align}
\Pi^{(n)}(q^2) = \frac{q^2}{2\pi i} \oint ds \, 
\frac{\Pi^{(n)}(s)}{s\,(s-q^2)} \,.
\label{Polfunc}
\end{align}
The integration contour in the variable $E$ (with $s=(2m+E)^2$) is drawn 
in Figure~\ref{fig:intcircuit}.  
For the terms relevant in this paper the functions $\Pi^{(n)}(s)$ have a 
two-particle cut starting at $s=4m^2$, equivalently $E=0$. In order 
to avoid touching the singular point at $E=0$, we separate an infinitesimal 
circle $C_{\varepsilon}$ of radius $\varepsilon$, parametrized as 
$E=\varepsilon\,e^{i\varphi}$, $\varphi\in (0,2\pi)$ from the remainder 
of the integration contour. The straight lines above and below the 
cut extend from $\varepsilon$ to $+\infty$ and 
involve the difference $\Pi(s+i\eta)-\Pi(s-i\eta)=2i\,{\rm Im}\,\Pi(s+i\eta)$
with $\eta$ positive-infinitesimal. The circle at infinity does not 
contribute to the once-subtracted dispersion relation, hence (\ref{Polfunc}) 
can be written as 
\begin{eqnarray}
\Pi^{(n)}(q^2) &=& \frac{q^2}{2 \pi i} \left\{ 
\Pi^{(n)}_{C_\varepsilon}(q^2) + \Pi^{(n)}_{\rm cont}(q^2) \right\}
\nonumber\\
&=& %\hspace*{-1.5cm} =
 \frac{q^2}{\pi i}
\int_{C_\varepsilon} \frac{dE}{2m+E} \, \frac{\Pi^{(n)}(E)}{(2m+E)^2-q^2} 
+ \frac{2q^2}{\pi }\int_{\varepsilon}^{\infty} \!\frac{dE}{2m+E} \, 
\frac{{\rm Im}\,\Pi^{(n)}(E+i\eta)}{(2m+E)^2-q^2}\,.\qquad 
\label{disprelE}
\end{eqnarray}
%
%\begin{align}
% \Pi^{(n)}_{C_\varepsilon}(q^2) & = 
%\int_{C_\varepsilon} \! dE \, \frac{\Pi^{(n)}(E)}{(2m+E)\,((2m+E)^2-q^2)} 
%\nonumber\\
%\Pi^{(n)}_{\rm cont}(q^2) & = 
%\int_{\varepsilon}^{\infty} \!\! dE \, \frac{2i\,{\rm Im}\,\Pi^{(n)}(E+i\eta)}{(2m+E)\,((2m+E)^2-q^2)} 
%\; ,
%\label{Piparts}
%\end{align}
%
The contribution from the small circle $C_\varepsilon$ vanishes for 
$\varepsilon\to 0$, if the vacuum polarization is less singular than 
$1/E$ at $E=0$; however, this condition is not satisfied in general. 
For the computation of the small-circle contribution 
$\Pi^{(n)}_{C_\varepsilon}(q^2)$ 
we can use the expansion 
\begin{align}
 \Pi^{(n)}(E) & = 
\Pi^{(n)}_{0}(E)  + \Pi^{(n)}_{1}(E) +\dots
\label{Pinonres}
\end{align}
of $\Pi^{(n)}(E)$ around threshold. The leading term, $\Pi^{(n)}_0(E)$, 
is given by the expansion in $\alpha$ of the zero-distance Coulomb Green 
function $G_0(0,0;E)$, see (\ref{eq:PiGrelation}), (\ref{eq:G0MSbar}) 
and (\ref{eq:Pipert}). For $n\ge 3$
we have
\begin{align}
\Pi_0^{(n)}(E) = \frac{\alpha^2}{2} \,\lambda^{n-2} \, \zeta_{n-1} 
\qquad (\lambda=\alpha/(2\sqrt{-E/m}))\,,
\label{PiLO}
\end{align}
from the expansion of the Digamma function $\Psi(1-\lambda)$. The 
next-to-leading term  $\Pi^{(n)}_1(E)$ is suppressed by $\sqrt{E/m}$, 
and equals the leading-order one times an ${\cal O}(\alpha)$ hard matching 
coefficient~\cite{Karplus:1952wp}:\footnote{The simplicity of this 
result is specific to the case of electrons, in which case the 
Coulomb potential receives no radiative corrections.}
\begin{align}
\Pi^{(n)}_{1}(E) & = -\frac{4 \alpha}{\pi}\, \Pi^{(n-1)}_{0}(E)  \, .
\label{PiNLO}
\end{align}
Similarly, $\Pi^{(n)}_2(E)$ can be extracted from the non-relativistic 
expansion of vacuum polarization at next-to-next-to-leading order (NNLO), 
and so on. 

It is clear that by construction the two integrals in (\ref{disprelE}) 
are well-defined, but each is singular for small $\varepsilon$. We now 
show explicitly that the sum is well-defined in the limit 
$\varepsilon\to 0$. Because of the relation (\ref{PiNLO}) it is 
sufficient to prove this for the leading  term
$\Pi^{(n)}_{0}(E)$ at the orders in $\alpha$ relevant to this paper. 
Once $\Pi^{(n)}_{2}(E)$ is included, integrals logarithmic in energy 
appear, but the generalization of the considerations below to this 
case is straightforward. Since we are only interested in the region 
$E\sim \varepsilon \to 0$, we can expand the denominators in the
integrands of (\ref{disprelE}) in $E/m$ and $mE/(4m^2-q^2)$. Then we have 
to prove that in the limit $\varepsilon\to 0$ 
\begin{align}
I_{C_\varepsilon}^{(n,k)}(\varepsilon) + I^{(n,k)}_{\rm cont}(\varepsilon)= 
{\cal O}(\varepsilon^0) \;,
\end{align}
where 
\begin{align}
I_{C_\varepsilon}^{(n,k)} (\varepsilon) & \equiv 
\frac{1}{2 i}\int_{C_\varepsilon} \!\! dE  \, E^k\,\Pi^{(n)}_0(E)\,,
\\[2mm]
I^{(n,k)}_{\rm cont}(\varepsilon) & \equiv 
\int_{\varepsilon}^{E_{\rm max}}\!\! dE \, E^k\, {\rm Im}\,\Pi^{(n)}_0(E) \,,
\label{Iints}
\end{align}
and $k=0,1,2\dots$. We have limited the integral along the cut up to a maximum energy $E_{\rm max}$ 
because the expansion in $E$ potentially leads to integrands which do not converge at infinity; this is
however irrelevant for the $\varepsilon \to0$ limit studied here.  

Given (\ref{PiLO}) the evaluation of the two integrals is straightforward 
and we find ($n\geq 3$)
\begin{align}
I_{C_\varepsilon}^{(n,k)} (\varepsilon) & =
\left\{\begin{array}{ll}
\displaystyle
  \frac{\alpha^n}{2^{n-1}} \, \zeta_{n-1} \,m^{k+1}\,
\frac{\sin \frac{n\pi}{2}}{n/2 -2- k} 
\,\Big(\frac{\varepsilon}{m}\Big)^{k+2-n/2}& \qquad k\not = \frac{n}{2}-2
\\[0.5cm]
\displaystyle
 -\frac{\pi \alpha^{n}}{2^{n-1}} \, \zeta_{n-1} 
\,(-m)^{n/2-1}& \qquad k = \frac{n}{2}-2
\end{array}
\right.
\label{ICeps}
\end{align}
and 
\begin{align}
I^{(n,k)}_{\rm cont}(\varepsilon) & =
- \frac{\alpha^n}{2^{n-1}} \, \zeta_{n-1}\, m^{k+1}\,
\frac{\sin \frac{n\pi}{2}}{n/2-2-k} 
\left[ \Big(\frac{\varepsilon}{m}\Big)^{k+2-n/2} \!\!\! -\Big(\frac{E_{\rm max}}{m}\Big)^{k+2-n/2} \right]\,.
\label{Icont}
\end{align}
Although we have chosen a particular form for the integration contour  
$C_{\varepsilon}$ surrounding $E=0$, we would get the same result 
(\ref{ICeps}) for any contour which has $E=\varepsilon\,e^{i2\pi^-}$
and $E=\varepsilon\,e^{i0^+}$ as initial and final points, respectively. 
Eqs.~(\ref{ICeps}) and~(\ref{Icont}) explicitly show the cancellation of 
the $\varepsilon$-divergent terms between integrals 
$I_{C_\varepsilon}^{(n,k)} (\varepsilon)$ and 
$I^{(n,k)}_{\rm cont}(\varepsilon)$ in the dispersion relation. It is 
also worth noting that (\ref{Icont}) vanishes for even $n$ and 
so does (\ref{ICeps}), except for the special case $k=n/2-2$ in which 
there is {\em only} the contribution (\ref{ICeps}) from the small circle.

We have thus shown that the dispersion relation (\ref{disprel}) must 
be modified in the presence of threshold singularities. The correct 
dispersion relation is threshold-subtracted and reads
\begin{align}
 \Pi^{(n)}(q^2) \; = \; 
\frac{q^2}{2 \pi i} \,
\Pi^{(n)}_{C_\varepsilon}(q^2)
+\frac{q^2}{\pi } 
\int_{(2m+\varepsilon)^2}^{\infty} \!\! ds \,
\frac{ {\rm Im}\,\Pi^{(n)}(s+i\eta)}{s\,(s-q^2)} 
\, .
\label{disprelgen}
\end{align}
In the 
following we consider the four- and five-loop case explicitly.

The four-loop case is directly related to the discussion in 
section~\ref{sec:equivalence}. For $n=4$ we must have $k=0$, and 
with 
\begin{align}
I_{C_\varepsilon}^{(4,0)} (\varepsilon) = \frac{\pi\alpha^4}{8}\,\zeta_3 \,m
\,,
\label{ICeps4}
\end{align}
the dispersion relation (\ref{disprelgen}) for the ${\cal O}(\alpha^4)$ 
vacuum polarization is
\begin{align}
\Pi^{(4)}(q^2)  = 
\frac{q^2}{8\,(4m^2-q^2)} \,\alpha^4\zeta_3\,
 + \frac{q^2}{\pi}\,
\int_{(2m+\varepsilon)^2}^{\infty} \!\! ds \, 
\frac{ {\rm Im}\,\Pi^{(4)}(s+i\eta)}{s\,(s-q^2)} 
\; ,
\label{Pi4}
\end{align}
where we have written back the continuum integral in terms of $s$.
The first term in~(\ref{Pi4}) reproduces the $\alpha^4/E$ term in $\Pi_0(E)$, 
see~(\ref{eq:Pipert}), if we specify $q^2=(2m+E)^2$.
In section~\ref{sec:equivalence} we interpreted this term in the distribution 
sense to extract its imaginary part, which then contributes to the electron 
anomalous moment. This contribution can now be understood as arising from 
an additional term in the dispersion relation obeyed by the vacuum 
polarization (see following section). Note that the integration over the 
continuum starts at $s=4m^2 +{\cal O}(\varepsilon)$. This prevents that the  
contribution from the $\alpha^4/E$ term in $\Pi^{(4)}(E)$ could
be double-counted by including its imaginary part localized at $E=0$ 
in the spectral density in the continuum integral.

On the other hand, for perturbative contributions to the spectral density 
with odd $n$, the contour integral around the threshold can be
interpreted as providing the necessary subtraction terms to regulate 
the divergence in ${\rm Im}\,\Pi_0^{(n)}$ when $E\to 0$. Explicitly, at 
the five-loop order, where the first divergence is found, 
since $\Pi_0^{(5)}(E) \sim \alpha^5 \, (-E/m)^{-3/2}$, the 
contribution from this term to the contour
$C_\varepsilon$ reads
\begin{align}
I_{C_\varepsilon}^{(5,k)} (\varepsilon) =  
\frac{\alpha^5}{8} \,\zeta_4\, m^{k+1}\,
\Big(\frac{\varepsilon}{m}\Big)^{k-1/2} \,\frac{1}{1-2k}\,,
\label{eq:IC5k}
\end{align}
which is only divergent for $k=0$. At ${\cal O}(\alpha^5)$ the NLO 
non-relativistic vacuum polarization $\Pi_1(E)$ is divergent
at $E=0$ for the first time, and thus also contributes to the 
$C_\varepsilon$-contour integral. The corresponding result is 
simply $(-4\alpha/\pi) I_{C_\varepsilon}^{(4,0)} (\varepsilon)$, 
see~(\ref{PiNLO}). Plugging this together with~(\ref{eq:IC5k}) for $k=0$ 
into the dispersion relation (\ref{disprelE}) or (\ref{disprelgen}),
we obtain
\begin{align}
 \Pi^{(5)}(q^2)  = & \
\frac{q^2}{2\,(4m^2-q^2)} \,\frac{\alpha^5}{\pi}\,
\left(  \frac{\zeta_4}{4} \,
\Big(\frac{\varepsilon}{m}\Big)^{-1/2}-  \, \zeta_3 \right)
\nonumber\\[2mm]
&  
+ \frac{q^2}{\pi}\,
\int_{(2m+\varepsilon)^2}^{\infty} \!\! ds \, \frac{ {\rm Im}\,\Pi^{(5)}(s+i\eta)}{s\,(s-q^2)} 
\bigg\}
\; ,
\label{Pi5}
\end{align}
which, using 
$ (\varepsilon/m)^{-1/2}=1/2 \int_\varepsilon^{\infty} dE/m \,(E/m)^{-3/2}$,
can be rewritten as
\begin{eqnarray}
 \Pi^{(5)}(q^2)  &=&
-\frac{q^2}{4 m^2-q^2} \,
 \frac{\alpha^5}{2\pi} \, \zeta_3
\nonumber \\[2mm] 
&& \hspace*{-1.5cm} +\,\frac{q^2}{\pi } 
\int_{(2m+\varepsilon)^2}^{\infty}  \frac{ds}{s} \,
\bigg\{ 
\frac{ {\rm Im}\,\Pi^{(5)}(s+i\eta)}{(s-q^2)} 
+
\frac{\alpha^5\,\zeta_4}{32\,(4m^2-q^2)} \,
\frac{\sqrt{s}}{m}\,
\Big( \frac{m }{ \sqrt{s}-2m } \Big)^{3/2} 
\bigg\}\,.
\label{Pi5subtract}
\end{eqnarray}
In this form of the integration over the spectral density
is well-defined at the $e^+e^-$ threshold. The second term in curly brackets 
effectively acts as a subtraction of the divergent behaviour of the 
first at $s=4m^2$, and the $\varepsilon$ in the integration boundary is 
only required as a reminder that the threshold-localized $\delta(E)$ 
term in ${\rm Im}\,\Pi^{(5)}(s)$ should not be included.

%%%%%%%%%%%%%%%%%%%%%%%%%%%%%%%%%%%%%%%%%%%%%%%%%%%%%%%%%%%%%%%%%%%%%%%%%%%%%
\section{\boldmath 
Dispersive representation of $a_e^{\rm (vp)}$ beyond 
${\cal O}(\alpha^4)$}
\label{sec:aedisp}

In this section we use the result from above to provide the corrected 
dispersive representation (\ref{dispaevp}) for the vacuum polarization 
contribution to the electron anomalous magnetic moment. 

At ${\cal O}(\alpha^5)$ and ${\cal O}(\alpha^6)$\footnote{No correction 
is required in lower orders as should be clear from the foregoing.}  
we set $q^2=-x^2 m^2/(1-x^2)$ and insert the dispersion 
relations (\ref{Pi4}), (\ref{Pi5}) for $\Pi^{(4)}(q^2)$ and 
$\Pi^{(5)}(q^2)$, respectively, into (\ref{aevp}). The 
results read
\begin{align}
a_e^{{\rm (vp)},(5)} & =  
\frac{\alpha^5}{8\pi} \,\zeta_3\,K(4m^2)
+ \frac{\alpha}{\pi^2} \int_{(2m+\varepsilon)^2}^\infty \frac{ds}{s} 
\,{\rm Im}\,\Pi^{\rm (4)}(s+i\eta)\,K(s)
\label{dispaevp5}
\end{align}
and 
\begin{eqnarray}
a_e^{{\rm (vp)},(6)} & =&   
-\frac{\alpha^6}{2\pi^2} \,\zeta_3\,K(4m^2)
\nonumber\\[2mm]
&& \hspace*{-1.5cm}+\, 
\frac{\alpha}{\pi^2} \int_{(2m+\varepsilon)^2}^\infty \frac{ds}{s} \,
\bigg\{ {\rm Im}\,\Pi^{\rm (5)}(s+i\eta)\,K(s) 
+ \frac{\alpha^5\zeta_4}{32}\, 
\frac{\sqrt{s}}{m}\,
\Big( \frac{m }{ \sqrt{s}-2m } \Big)^{3/2}\,K(4m^2)\,\bigg\}\,.
\quad
\label{dispaevp6}
\end{eqnarray}
These two equations provide the correct expressions for the computation of
the ${\cal O}(\alpha^5)$  and (currently unknown)  ${\cal O}(\alpha^6)$ 
corrections to the electron anomalous magnetic moment induced by  
the ${\cal O}(\alpha^4)$ and ${\cal O}(\alpha^5)$ vacuum polarization 
insertions, respectively, exploiting perturbative approximations to the 
spectral density from $e^+e^-$ intermediate states\footnote{Recall that 
the spectral density from intermediate three-photon states, which start 
to contribute to the vacuum polarization at ${\cal O}(\alpha^4)$, 
has to be added separately to the formulae above.} without any 
resummation.  In particular, the dispersive representation for
$a_e^{{\rm (vp)},(6)}$ above is now suitable for numerical integration, 
since the singular $1/(\sqrt{s}-2m)^{3/2}$ behaviour of 
${\rm Im}\,\Pi^{\rm (5)}$ at threshold gets cancelled by the second term. 
The small $\varepsilon$ dependence in the lower integration limit serves as 
a reminder that no imaginary part of the form $\delta(E)$ should be 
accounted for in the spectral density.

The dispersive representation at even higher orders in $\alpha$ can 
obtained in a similar way. At ${\cal O}(\alpha^7)$ one has to consider 
terms arising from the NNLO non-relativistic vacuum polarization
$\Pi_2(E)$  in the computation of $\Pi^{(n)}_{C_\varepsilon}(q^2)$. 
Given that the calculation of the ${\cal O}(\alpha^6)$ electron anomalous 
magnetic moment has not yet been attempted, it is unlikely that the 
expression for $a_e^{{\rm (vp)},(7)}$ would be needed in the foreseeable 
future, and we do not pursue this order further here. 

%%%%%%%%%%%%%%%%%%%%%%%%%%%%%%%%%%%%%%%%%%%%%%%%%%%%%%%%%%%%%%%%%%%%%%%%%%%
\begin{figure}[t]
\begin{center}
\includegraphics[scale=0.75]{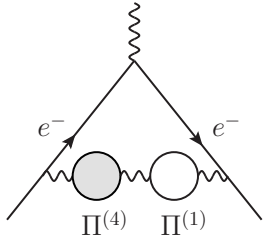}
\caption{The relevant  ${\cal O}(\alpha^6)$ contribution with two vacuum 
polarization insertions. The symmetric diagram must also be considered.}
\label{fig:2vpdiag}
\end{center}
\end{figure}
%%%%%%%%%%%%%%%%%%%%%%%%%%%%%%%%%%%%%%%%%%%%%%%%%%%%%%%%%%%%%%%%%%%%%%%%%%%

Let us finally mention that at ${\cal O}(\alpha^6)$ the dispersive 
representation of the diagrams with two vacuum polarization insertions 
also receives an additional contribution. It is easily obtained from the 
resummed version of~(\ref{aevp}) (see, for instance, Eq.~(70) 
of~\cite{Jegerlehner:2009ry}). Retaining the
relevant term $2\,\Pi^{(4)}\, \Pi^{(1)}$, see Figure~\ref{fig:2vpdiag}, 
we have
\begin{align}
\delta a_e^{{\rm (vp)},(6)} & = \frac{2\alpha}{\pi} \int_0^1 dx \,(1-x) \, 
\Pi^{(4)}\Big( \frac{-x^2}{1-x}\, m^2 \Big) 
\,
\Pi^{(1)}\Big( \frac{-x^2}{1-x}\, m^2 \Big)
\;.
\label{ae2vp}
\end{align}
Inserting the dispersion relation~(\ref{Pi4}) for $\Pi^{(4)}$ and 
treating $\Pi^{(1)}$ as part of a new kernel function, we obtain 
\begin{align}
\delta a_e^{{\rm (vp)},(6)}  = &  
-\frac{\alpha^5}{4\pi} \,\zeta_3\,K^{(1)}(4m^2)
-\frac{2\alpha}{\pi^2} \int_{(2m+\varepsilon)^2}^\infty \frac{ds}{s} 
\,{\rm Im}\,\Pi^{\rm (4)}(s+i\eta)\,K^{(1)}(s)\,,
\label{dispae2vp5}
\end{align}
with
\begin{align}
K^{(1)}(s) & \equiv \int_0^1 dx \, \frac{x^2(1-x)}{x^2+(1-x)s/m^2} 
\,\Pi^{(1)}\Big( \frac{-x^2}{1-x}\, m^2 \Big)\,.
\label{Ktilde}
\end{align}
The analytic expression for the one-loop vacuum  
polarization $\Pi^{(1)}(q^2)$ can
be found, for instance, in~\cite{Jegerlehner:2009ry}.

%%%%%%%%%%%%%%%%%%%%%%%%%%%%%%%%%%%%%%%%%%%%%%%%%%%%%%%%%%%%%%%%%%%%%%%%%%
\section{Hadronic pair production}
\label{sec:ttbar}

Threshold singularities are also present in higher-order perturbative 
calculations of heavy particle pair production cross sections. The 
analysis of dispersion relations for the photon vacuum polarization 
provides the clue to solving a related divergence problem in the 
computation of the total pair production cross section. The following 
applies to any particle species (for instance, of supersymmetric particles), 
but we discuss it for the specific case of top quark production in hadron 
collisions, which is 
presently the most relevant one. The generalization should be evident.

The total hadronic cross section for the production of a $t\bar t+X$ 
final state in collisions of hadrons $N_{1,2}$ with centre-of-mass (cms) 
energy $s$ is obtained from
\begin{eqnarray}
\sigma_{N_1 N_2\to t\bar t X}(s) &=&
\sum_{p,p'=q,\bar q,g}\,\int_{4 m_t^2/s}^1 \!d\tau\,L_{pp'}(\tau,\mu_f)
\,\hat\sigma_{pp'} (s \tau,\mu_f)
\nonumber\\
&=& \sum_{p,p'=q,\bar q,g}\,\int_0^{\sqrt{1-4 m_t^2/s}} \!\!d\beta\,
L_{pp'}(\beta,\mu_f)\,\frac{8\beta m_{t}^2}{s(1-\beta^2)^2}
\,\hat\sigma_{pp'}(\beta,\mu_f) \, .
\label{eq:sig-had}
\end{eqnarray}
Here $\hat\sigma_{pp'} (s \tau,\mu_f)$ is the (factorization-scale dependent) 
partonic cross section for partonic cms energy $\hat s = \tau s$, 
$\beta = \sqrt{1-4 m_t^2/\hat s}$, and 
the parton luminosity is defined in terms of the parton distributions 
functions~(PDFs) via
\begin{equation}
\label{eq:lumi}
L_{p p^\prime}(\tau,\mu) = \int_0^1 dx_1
dx_2\,\delta(x_1 x_2 - \tau) \,f_{p/N_1}(x_1,\mu)f_{p^\prime/N_2}(x_2,\mu)\,.
\end{equation}
The parton luminosity approaches a constant near threshold 
$\hat s\to 4 m_t^2$, equivalently $\beta\to 0$. The most singular 
behaviour of the partonic cross section is $\alpha_s^2/m_t^2 \times 
\beta \times (\alpha_s/\beta)^k$, where in common terminology 
$k=1$ refers to the next-to-leading order correction to the cross section, 
$k=2$ to NNLO, and so on. The leading behaviour is absent for 
$k=3$, where instead it is given by $\alpha_s^2/m_t^2 \times 
\beta \times (\alpha_s/\beta)^2\times \alpha_s 
\ln^2\beta$~\cite{Beneke:2011mq}. 
Hence the convolution with the parton luminosity diverges beginning 
at order ${\cal O}(\alpha_s^6)$ or N4LO. On the other hand, when the 
singular terms are summed into the Coulomb Green function, the 
convolution becomes convergent and the net effect of the Coulomb 
corrections is very small for the total cross section. These facts 
were noted in \cite{Beneke:2011mq}, but lead to a puzzling situation. 
Resummation should not be required to compute a small effect, or 
make the total cross section well-defined. Rather, conventional 
fixed-order perturbation theory should provide the correct result 
directly. 

To approach the problem, we note that the partonic cross sections 
can be related to the discontinuity of the forward parton scattering 
amplitude
\begin{equation}
\hat\sigma_{pp'}(\hat s) = \frac{1}{\hat s}\,\mbox{Im}_{\,t\bar t}\,
{\cal A}(pp'\to pp')(\hat s)\,.
\label{eq:forward}
\end{equation}
We then observe the similarity of the first line of (\ref{eq:sig-had}) 
and the dispersive representation (\ref{dispaevp}) of the vacuum 
polarization to the electron anomalous magnetic moment, if we identify the 
forward amplitude ${\cal A}_{pp'}(\tau s)\equiv 
{\cal A}(pp'\to pp')(\tau s)$ with the vacuum polarization 
$\Pi(s)$, and the parton luminosity $L_{pp'}(\tau,\mu_f)$ 
with the kernel function $K(s)$. The subscript ``$t\bar t\,$'' in 
(\ref{eq:forward}) means that only the cuts with a top-antitop pair 
should be included. We can ignore the other cuts, since they do 
not produce threshold singularities at $4 m_t^2$, in the same way as 
the three-photon intermediate states that contribute to the photon vacuum 
polarization were of no relevance to the previous discussion.

The correspondence makes it clear, how the convolution (\ref{eq:sig-had}) 
should be defined, at each order in perturbation theory, 
when the parton cross section develops non-integrable 
threshold singularities. We first write down the dispersion relation 
for ${\cal A}_{pp'}(\hat s)$ with a circle of infinitesimal 
radius around $4 m_t^2$ separated, exactly as in (\ref{disprelgen}). 
The integral term in this relation leads to (\ref{eq:sig-had}) with 
the lower limit modified to $(2 m_t+\epsilon)^2/s$ and a corresponding 
adjustment of the second line. The contribution 
${\cal A}^{(n)}_{C_\varepsilon,\,pp'}(\hat s)$ from the small 
circle has to be added as an extra contribution to the hadronic 
cross section. Since both terms in the subtracted dispersion 
relation (\ref{disprelgen}) are separately divergent as $\epsilon\to 0$, 
it is again convenient to rewrite the circle contribution as a 
subtraction in the integrand of the cut contribution, similar to 
(\ref{Pi5}), (\ref{dispaevp6}). When this is done, only the 
threshold-localized terms in the spectral function/imaginary part of 
the forward amplitude remain to be added explicitly. In other words, 
the correct modification of (\ref{eq:sig-had}) implies subtracting the 
partonic cross sections appropriately and adding the delta-function 
contributions.

The subtraction terms can be determined from the expansion of the 
forward scattering amplitude near the top threshold. It is convenient 
to split the production cross section into contributions from 
$t\bar t$ states in a given irreducible colour representation $R_\alpha$. 
Near threshold, the amplitude can be written in the form 
\begin{equation}
i {\cal A}_{pp'}^{R_\alpha}=\int d^4 x\,
\braket{p p'|\Tprod{i {\cal O}_{pp'}^{R_\alpha\dagger }(0)i
{\cal O}_{pp'}^{R_\alpha}(x)}|p p'}.
\end{equation}
where ${\cal O}_{pp'}^{R_\alpha}$ is a local operator, which produces 
a $t\bar t$ pair in representation $R_\alpha$ from the $p p'$ parton 
initial state \cite{Beneke:2011mq}. The leading term in the 
threshold expansion (similar to (\ref{Pinonres})) to all orders in 
perturbation theory reads
\begin{equation}
\frac{1}{\hat s}\,{\cal A}_{pp',0}^{R_\alpha}(\hat s) = 
\frac{4\pi^2 \alpha_s^2}{m_t^4}\,\sigma_{pp'}^{R_\alpha}\,
G_0^{R_\alpha}(0,0;E)
\label{eq:forwardthres}
\end{equation}
where $G_0^{R_\alpha}(0,0;E)$ is the Coulomb Green function for the 
colour representation $R_\alpha$,  given by (\ref{eq:G0MSbar}) with 
$\alpha \to -\alpha_s D_{R_\alpha}$. The relevant cases are the 
attractive colour-singlet channel, $D_1 = - C_F = -4/3$, and the 
repulsive octet one, $D_8=1/(2 N_c) = 1/6$.\footnote{
Although not relevant
for the following, it is instructive to see how the equivalence of
the ``non-perturbative'', resummed calculation and the fixed-order one
discussed in Section~\ref{sec:equivalence} works for a
repulsive Coulomb force ($\alpha<0$).
The imaginary part (\ref{eq:ImG0}) of the
resummed spectral function has no bound-state contribution in this
case, while the expression for the Sommerfeld continuum remains
unchanged. However, the velocity integral in (\ref{eq:contint})
and in the footnote there is proportional to $-\alpha |\alpha|$, and 
changes sign for negative $\alpha$. 
As a consequence the first term in (\ref{eq:resummed2}) is absent, while
the second changes sign (that is, equals, $+\alpha^5\zeta_3/(8\pi)\times
K(4 m^2)$), which yields again agreement with the
perturbative result (\ref{eq:F)}).}
The constants 
$\sigma_{pp'}^{R_\alpha}$ can be found by comparison with the
threshold-limit of the Born cross section to be
\begin{equation}
\sigma_{gg}^{1} = \frac{1}{96}\,,
\quad\qquad 
\sigma_{gg}^{8} = \frac{5}{192}\,,
\quad\qquad 
\sigma_{q\bar q}^{1} = 0\,,
\quad\qquad 
\sigma_{q\bar q}^{8} = \frac{1}{9}\,.
\end{equation}
The perturbative expansion of (\ref{eq:forwardthres}) is very similar to
(\ref{PiLO}).

The inclusive top pair production cross section is presently known to 
${\cal O}(\alpha_s^4)$ or NNLO in perturbation theory \cite{Czakon:2013goa}. 
It is therefore of particular interest to investigate the implications 
of the above discussed modification of (\ref{eq:sig-had}) at the 
next order, where indeed it arises for the first time. We recall that 
at this order there is no explicit divergence of the convolution integral, 
but since
\begin{equation}
\frac{1}{\hat s}\,{\cal A}_{pp',0}^{R_\alpha,(5)}(\hat s) = 
-\frac{\pi \alpha_s^2}{4 m_t^2}\,\sigma_{pp'}^{R_\alpha}\,
(-\alpha_s D_{R_{\alpha}})^3\,
\frac{\zeta_3}{E/m_t}
\label{eq:forward5}
\end{equation}
causes a threshold-localized term $\delta(E)$, the threshold-subtracted 
dispersion relation contains a non-vanishing contribution from the circle, 
as in (\ref{Pi4}), (\ref{dispaevp5}). The result therefore reads 
\begin{eqnarray}
\sigma_{N_1 N_2\to t\bar t X}^{\rm N3LO}(s) &=&
\frac{\pi^2\zeta_3\alpha_s^5}{s} 
\sum_{pp'=q\bar q,gg}\sum_{R_{\alpha}=1,8}\,(-D_{R_\alpha})^3 
\sigma_{pp'}^{R_\alpha}\,L_{pp'}(4 m_t^2/s,\mu_f)
\nonumber\\
&& +\,\sum_{p,p'=q,\bar q,g}\,\int_{(2m_t+\varepsilon)^2/s}^1 
\!d\tau\,L_{pp'}(\tau,\mu_f)
\,\hat\sigma_{pp'}^{\rm N3LO}(s \tau,\mu_f)\, .
\label{eq:sig-hadmod}
\end{eqnarray}
If it ever becomes feasible to compute the N3LO partonic cross section 
$\hat\sigma_{pp'}^{\rm N3LO}(\hat s)$, it will most likely be as a sum 
of virtual and real contributions, integrated numerically over phase space, 
as presently done at NNLO \cite{Czakon:2013goa}. In this case, the 
most singular behaviour would be found to be $1/\beta^2\times \ln^2\beta$ 
\cite{Beneke:2011mq}, but the delta-function contribution would be 
missed. The term in the first line of the previous 
equation must be added explicitly to such a computation. Numerically, 
however, this additional contribution is very small, as shown in 
Table~\ref{tab:result}. This amounts to about or less than a per mil of 
the total top pair production cross section, and is about an order of 
magnitude smaller 
than the cross section beyond NNLO due to the next-to-next-to-leading 
logarithmic (NNLL) resummation of 
Coulomb and soft emission effects~\cite{Beneke:2012wb}.\footnote{
Note that the first line of (\ref{eq:sig-hadmod}) {\em is} included 
in~\cite{Beneke:2012wb}, since the resummed partonic cross section 
includes the bound state poles and the Sommerfeld continuum, amounting 
to the ``non-perturbative'' computation in the terminology employed 
here.} 

The reason for the smallness of the additional contributions is that 
top pairs are predominantly produced in the colour-octet state, but the 
octet contribution to the first line of (\ref{eq:sig-hadmod}) is 
suppressed by $(D_8/D_1)^3=-1/512$ due to the small colour factor. If 
a new species of heavy strongly interacting particles were produced in 
a singlet state or another colour state with a strong Coulomb interaction, 
no matter whether attractive or repulsive, the threshold-localized 
term could make a relevant contribution to the total cross section.
 
%%%%%%%%%%%%%%%%%%%%%%%%%%%%%%%%%%%%%%%%%%%%%%%%%%%%%%%%%%%%%%%%%%%%%%%%%%
\begin{table}[t]
\newcommand{\m}{\hphantom{$-$}}
\newcommand{\cc}[1]{\multicolumn{1}{c}{#1}}
\renewcommand{\tabcolsep}{0.8pc} % enlarge column spacing
\renewcommand{\arraystretch}{1.0} % enlarge line spacing
\begin{center}
\begin{tabular}{@{}cccccc}
\hline  \vspace{-4mm} \\  
     & \hspace{-4mm} \mbox{Tevatron} 
     & \hspace{-3mm} LHC ($7\,$TeV) 
     & \hspace{-3mm} LHC ($8\,$TeV) 
     & \hspace{-3mm} LHC ($13\,$TeV) 
     & \hspace{-3mm} LHC ($14\,$TeV) \\
\hline   \\[-2mm]
\hspace{-7mm}
\phantom{ab}  
     & \hspace{-5mm} $\phantom{0}0.0016$
     & \hspace{-5mm} $\phantom{0}0.14$
     & \hspace{-5mm} $\phantom{0}0.19$
     & \hspace{-5mm} $\phantom{0}0.59$
     & \hspace{-5mm} $\phantom{0}0.69$
 \vspace{2mm} \\
\hline
\end{tabular}\\[2pt]
\end{center}
\caption{Additional threshold-localized contribution to the 
inclusive N3LO top quark pair production cross section for 
the Tevatron $p\bar p$ and LHC $pp$ collider at various cms energies as 
given in brackets. The MSTW NNLO PDFs \cite{Martin:2009iq} with 
$\mu_f=m_t=173.3\,$GeV and $\alpha_s=\alpha_s(m_t)=0.1085$ have been 
used in the evaluation. All cross sections in pb.}
\label{tab:result}
\end{table} 
%%%%%%%%%%%%%%%%%%%%%%%%%%%%%%%%%%%%%%%%%%%%%%%%%%%%%%%%%%%%%%%%%%%%%%%%%%

%%%%%%%%%%%%%%%%%%%%%%%%%%%%%%%%%%%%%%%%%%%%%%%%%%%%%%%%%%%%%%%%%%%%%%%%%%%%
\section{Summary}
\label{sec:conclusion}

Inspired by a recent controversy over whether the positronium pole 
contribution needs to be added explicitly to the dispersive representation 
of the vacuum polarization contribution to the electron magnetic 
moment, we showed how this contribution is accounted for in a 
direct fixed-order computation. This has led us to a more general 
consideration of dispersion relations in the presence of a pair 
particle production threshold. We find that the dispersion relation 
requires threshold subtractions, see (\ref{disprelgen}), similar to 
subtractions that are often required to account for the ultraviolet 
behaviour. The threshold-subtraction term can be determined from 
the expansion of vacuum polarization near the threshold. While our 
results imply that the dispersion relation receives additional terms, 
no correction of the anomalous magnetic moment is implied, since the 
evaluation in \cite{Aoyama:2010zp} is based on the integration of the 
vacuum polarization at Euclidean momenta. On the other hand, we 
find interesting ramifications for hadron-collider production of 
pairs of heavy particles, for which a Euclidean formulation is not 
available. When the computation is performed in the usual way as 
an integral of real and virtual corrections over phase space at a given 
order in the expansion in the strong coupling, 
an additional contribution has to be added at N3LO and the 
convolution of the partonic cross section with the parton luminosity 
must be modified from N4LO. We explicitly evaluated the N3LO contribution 
for hadronic top pair production and found that it is numerically 
small, of order of a per mil of the cross section.

\noindent
\subsubsection*{Acknowledgements}
We thank Antonio Pich for discussions on dispersion relations and Eric 
Laenen for comments on the manuscript. The work of MB is supported by the 
BMBF grant 05H15WOCAA. 
MB thanks the Kavli Institute for Theoretical Physics, Santa Barbara, for 
hospitality while this work was written up.

%%%%%%%%%%%%%%%%%%%%%%%%%%%%%%%%%%%%%%%%%%%%%%%%%%%%%%%%%%%%%%%%%%%%%%%%%%


\begin{thebibliography}{99}

\bibitem{Lautrup:1969uk}
  B.~E.~Lautrup,
  %``On sixth-order radiative corrections to the muon g-factor,''
  Nuovo Cim.\ A {\bf 64} (1969) 322.
  %%CITATION = NUCIA,A64,322;%%

\bibitem{Lautrup:1971jf}
  B.~E.~Lautrup, A.~Peterman and E.~de Rafael,
  %``Recent developments in the comparison between theory and experiments in quantum electrodynamics,''
  Phys.\ Rept.\  {\bf 3} (1972) 193.
  %%CITATION = PRPLC,3,193;%%

\bibitem{Mishima:2013ama}
  G.~Mishima,
  %``Bound State Effect on the Electron g-2,''
  arXiv:1311.7109 [hep-ph].
  %%CITATION = ARXIV:1311.7109;%%

\bibitem{Aoyama:2010zp}
  T.~Aoyama, M.~Hayakawa, T.~Kinoshita and M.~Nio,
  %``Proper Eighth-Order Vacuum-Polarization Function and its Contribution to the Tenth-Order Lepton g-2,''
  Phys.\ Rev.\ D {\bf 83} (2011) 053003,
  arXiv:1012.5569 [hep-ph].
  %%CITATION = ARXIV:1012.5569;%%

%\cite{Melnikov:2014lwa}
\bibitem{Melnikov:2014lwa}
  K.~Melnikov, A.~Vainshtein and M.~Voloshin,
  %``Remarks on the effect of bound states and threshold in g-2,''
  Phys.\ Rev.\ D {\bf 90} (2014) 017301,
  %doi:10.1103/PhysRevD.90.017301
  arXiv:1402.5690 [hep-ph].
  %%CITATION = doi:10.1103/PhysRevD.90.017301;%%

%\cite{Eides:2014swa}
\bibitem{Eides:2014swa}
  M.~I.~Eides,
  %``Recent ideas on the calculation of lepton anomalous magnetic moments,''
  Phys.\ Rev.\ D {\bf 90} (2014) 057301,
  %doi:10.1103/PhysRevD.90.057301
  arXiv:1402.5860 [hep-ph].
  %%CITATION = doi:10.1103/PhysRevD.90.057301;%%

%\cite{Hayakawa:2014tla}
\bibitem{Hayakawa:2014tla}
  M.~Hayakawa,
  %``Positronium resonance contribution to the electron g-2,''
  arXiv:1403.0416 [hep-ph].
  %%CITATION = ARXIV:1403.0416;%%

%\cite{Eiras:1999xx}
\bibitem{Eiras:1999xx}
  D.~Eiras and J.~Soto,
  %``Effective field theory approach to pionium,''
  Phys.\ Rev.\ D {\bf 61} (2000) 114027,
  %doi:10.1103/PhysRevD.61.114027
  [hep-ph/9905543].

\bibitem{Beneke:1999zr}
M.~Beneke, {\it {Perturbative heavy quark-antiquark systems}}, in: Proceedings of the 8th International Symposium on Heavy Flavor Physics (Heavy Flavors 8),  25-29 July 1999, Southampton, England, [hep-ph/9911490]

%\cite{Beneke:1997zp}
\bibitem{Beneke:1997zp}
  M.~Beneke and V.~A.~Smirnov,
  %``Asymptotic expansion of Feynman integrals near threshold,''
  Nucl.\ Phys.\ B {\bf 522} (1998) 321,
  %doi:10.1016/S0550-3213(98)00138-2
  [hep-ph/9711391].
  %%CITATION = doi:10.1016/S0550-3213(98)00138-2;%%

%\cite{Kiyo:2009gb}
\bibitem{Kiyo:2009gb}
  Y.~Kiyo, A.~Maier, P.~Maierhofer and P.~Marquard,
  %``Reconstruction of heavy quark current correlators at O(alpha(s)**3),''
  Nucl.\ Phys.\ B {\bf 823} (2009) 269,
  %doi:10.1016/j.nuclphysb.2009.08.010
  arXiv:0907.2120 [hep-ph].

%\cite{Karplus:1952wp}
\bibitem{Karplus:1952wp}
  R.~Karplus and A.~Klein,
  %``Electrodynamics displacement of atomic energy levels. 3. The Hyperfine structure of positronium,''
  Phys.\ Rev.\  {\bf 87} (1952) 848.
  %doi:10.1103/PhysRev.87.848
  %%CITATION = doi:10.1103/PhysRev.87.848;%%

\bibitem{Jegerlehner:2009ry}
  F.~Jegerlehner and A.~Nyffeler,
  %``The Muon g-2,''
  Phys.\ Rept.\  {\bf 477} (2009) 1,
  arXiv:0902.3360 [hep-ph].

%\cite{Beneke:2011mq}
\bibitem{Beneke:2011mq}
  M.~Beneke, P.~Falgari, S.~Klein and C.~Schwinn,
  %``Hadronic top-quark pair production with NNLL threshold resummation,''
  Nucl.\ Phys.\ B {\bf 855} (2012) 695,
  %doi:10.1016/j.nuclphysb.2011.10.021
  arXiv:1109.1536 [hep-ph].

%\cite{Czakon:2013goa}
\bibitem{Czakon:2013goa}
  M.~Czakon, P.~Fiedler and A.~Mitov,
  %``Total Top-Quark Pair-Production Cross Section at Hadron Colliders Through $O(~\frac{4}{S})$,''
  Phys.\ Rev.\ Lett.\  {\bf 110} (2013) 252004, 
  %doi:10.1103/PhysRevLett.110.252004
  arXiv:1303.6254 [hep-ph].

%\cite{Martin:2009iq}
\bibitem{Martin:2009iq}
  A.~D.~Martin, W.~J.~Stirling, R.~S.~Thorne and G.~Watt,
  %``Parton distributions for the LHC,''
  Eur.\ Phys.\ J.\ C {\bf 63} (2009) 189,
  %doi:10.1140/epjc/s10052-009-1072-5
  arXiv:0901.0002 [hep-ph].
  %%CITATION = doi:10.1140/epjc/s10052-009-1072-5;%%

\bibitem{Beneke:2012wb}
  M.~Beneke, P.~Falgari, S.~Klein, J.~Piclum, C.~Schwinn, M.~Ubiali and F.~Yan,
  %``Inclusive Top-Pair Production Phenomenology with TOPIXS,''
  JHEP {\bf 1207} (2012) 194, 
  %doi:10.1007/JHEP07(2012)194
  arXiv:1206.2454 [hep-ph].
 
%\bibitem{Fael:2014nha}
%  M.~Fael and M.~Passera,
  %``On the positronium contribution to the electron g-2,''
%  arXiv:1402.1575 [hep-ph].
  %%CITATION = ARXIV:1402.1575;%%

\end{thebibliography}
\end{document}